\documentclass[12pt]{elsarticle}
\usepackage{amsmath}
\usepackage{setspace}
\usepackage{graphicx}
\usepackage{tabularx}
\usepackage{amsthm}
\usepackage{amssymb}
\usepackage{lineno}
\usepackage{rotating}
\usepackage{epsfig}
\usepackage{epstopdf}
\usepackage{dcolumn}
\usepackage{bm}
\usepackage{mathptmx}
\usepackage[font=small]{caption}
\usepackage{setspace}
\usepackage{threeparttable}
\usepackage{booktabs}
\usepackage{tabularx}
\usepackage{float}
\usepackage[pagebackref=true]{hyperref}
\usepackage{graphicx}
\usepackage{chngcntr}
\usepackage[margin=1in]{geometry}

\begin{document}

\def\br{{\bf r}}
\def\et{{\em et al}}

\title{Machine-Learned Potential Energy Surfaces for Free Sodium Clusters with
Density Functional Accuracy: Applications to Melting}

\author[1]{Balasaheb J. Nagare\corref{cor1}%
  }
\ead{bjnagare@physics.mu.ac.in}

\author[1]{Sajeev Chacko}
\ead{sajeev.chacko@physics.mu.ac.in}

 \cortext[cor1]{Corresponding author}


 \affiliation[1]{organization={Department of Physics, University of Mumbai}, 
                 addressline={Kalina Campus, Santacruz (E)},
                 postcode={400 098},
                 city={Mumbai}, 
                 country={India}}

\author[3]{Dilip. G. Kanhere}
\ead{dgkanhere@gmail.com}

 \affiliation[3]{organization={Department of Scientific Computing, Modeling and Simulations, Savitribai Phule Pune University},
                 addressline={Ganeshkhind, Aundh}, 
                 city={Pune},
                 postcode={411 007}, 
                 country={India}}
\date{\today}

\begin{abstract}
    Gaussian Process Regression-based Gaussian Approximation Potential
    has been used to develop machine-learned interatomic
    potentials having density-functional accuracy
    for free sodium clusters.
    The training data was generated from a large sample of over 100,000 
    data points computed for clusters in the size range of $N=40-200$, 
    using the density-functional method as implemented in the VASP package.
    Two models have been developed, model M1 using data for $N$=55
    only, and model M2 using additional data from larger clusters.
    The models are intended for computing thermodynamic
    properties using molecular dynamics.
    Hence, particular attention has been paid to improve the fitting of the forces.
    Interestingly, it turns out that the best fit can be obtained by carefully 
    selecting a smaller number of data points {\em viz}. 1,900
    and 1,300 configurations, respectively, for the two models M1 and M2.
    Although it was possible to obtain a good fit using the data of Na$_{55}$ 
    only, additional data points from larger clusters were needed to get 
    better accuracies in energies and forces for larger sizes.
    Surprisingly, the model M1 could be significantly improved by 
    adding about 50 data points per cluster from the larger sizes.
    Both models have been deployed to compute heat capacities
    of Na$_{55}$ and Na$_{147}$ and to obtain about 40 isomers for 
    larger clusters of sizes $N=147, 200, 201$, and 252.
    There is an excellent agreement between the computed and 
    experimentally measured melting temperatures.
    The geometries of these isomers when further optimized by DFT, 
    the mean absolute error in the energies between DFT results and 
    those of our models is about 7 meV/atom or less.
    The errors in the interatomic bond lengths are estimated 
    to be below 2$\%$ in almost all the cases.
\end{abstract}

\maketitle


\section{Introduction}

    Molecular dynamics~(MD) has emerged as a powerful tool for
    investigating a wide variety of problems across different
    disciplines, including biology, chemistry, physics, material science and
    engineering~\cite{frenkel23, smith00, johnson05, vivo16, ciccoti22, gowthaman23}. 
    MD has gained popularity due to its conceptual simplicity and the 
    availability of highly mature and developed codes~\cite{plimpton95, 
    abraham15, gale97}.
    The method allows researchers to simulate the motion of atoms
    and molecules over time, providing valuable insights into the
    dynamics and properties of systems by an accurate and detailed description 
    at the microscopic level. By capturing atomic level details and dynamical
    behaviour, MD serves as a versatile and indispensable tool for understanding 
    complex systems. 
    
    The density functional theory (DFT) is a powerful electronic structure
    method that can provide an accurate description of many-electron systems.
    DFT-based molecular dynamics simulations, such as the Born-Oppenheimer
    or Car-Parrinello dynamics are very successful in 
    capturing the electronic structure and dynamics of materials~\cite{kohnoff,
    martin, car85}.
    The computational methods based on DFT are parameter-free 
    and able to capture the changing nature of 
    interatomic bonds as dynamics proceeds, on the fly.
    At every time step, the electrons in the system are maintained 
    on the Born-Oppenheimer surface. However, these methods are 
    computationally expensive and are not feasible for large systems 
    containing more than 1000 electrons or for long time scales,
    on a routine basis, especially with limited computational resources.

    To combine the advantages of classical MD, that is the speed and the accuracy
    of density-functional theory (DFT), machine learning approaches have
    emerged as a promising alternative for constructing highly accurate
    and computationally efficient potential energy surfaces (PES).
    During the last 15 years or so a number of techniques have been developed
    based on modern machine learning methods, and the field is still 
    continuously evolving~\cite{behler07, behler11, bartok10, nacl15, zhang18, 
    zuo20, pinheiro21, nov20, roh23}.
    The readers are referred to a number of excellent reviews~\cite{zuo20,
    pinheiro21, brown19}.
    One of the most significant works of Behler and Parrinello presented 
    a generalised neural network technique for modelling accurate PES~\cite{behler07,
    behler11, behler21}. 
    The method decomposes total energy $E$ as a sum of atomic contributions
    and uses atom-centred symmetry functions as descriptors.
    Another powerful technique developed by Bartok and Coworker 
    uses Gaussian Process Regression~(GPR)~\cite{bartok10}.
    It has been applied 
    to a number of extended systems~\cite{bartok13, bartok15}.
    Equivariant message passing network based 
    on the popular Atomic Cluster Expansion~(MACE)~\cite{Batatia2022mace} have been
    proposed by Batatia and coworkers. It utilizes graph networks to construct 
    highly accurate empirical potentials. It has been applied to a range 
    of systems including organic molecules, nanoparticles, and crystalline 
    materials~\cite{Batatia2022mace, Batatia2022Design}.
    Zhang {\em et. al.} developed the DeePMD method based on deep neural
    network~\cite{zhang18}. A number of successful applications have 
    been reported for reaction dynamics, catalytic processes, and 
    material properties~\cite{brown19, smith21}.
    In fact, there are a variety of networks and models reported 
    using different networks.
    The reader is referred to a number of reviews~\cite{roh23, chen22}.
    
    Most of the interatomic potentials developed by machine learning potential~(MLP) 
    are centred around extended systems either periodic or otherwise
    including homogeneous, heterogeneous and multicomponent 
    systems~\cite{zhang18, behler21, mishin21, unke21, wen22}.
    These applications demonstrated the efficiency of MLPs for computing 
    a variety of material properties. They also demonstrated the ability 
    to explore the phase  space more efficiently, allowing for the 
    identification of new phases and properties that were previously 
    unknown~\cite{seko20}.
    However, very few studies have been reported on finite-size systems, 
    viz., atomic clusters~\cite{smith21, sun19, chiriki16, wang22, mozh22, fronzi23}.
    Sun \et.~\cite{sun19} used the high-dimensional neural network potentials~(HDNNP) to 
    study the geometries of small platinum clusters in the size range of $N=6-20$. 
    Chiriki~\cite{chiriki16}~{\em et. al.} have used Monte Carlo~(MC) 
    simulations to study the thermodynamic properties of Na$_N$ 
    with $N=20-40$ using a neural network potential. 
    The reported error in energy was 20~meV/atom on the training data set.
    For Na$_{40}$, the main melting peak agreed with experiments.
    There has also been a development of moment tensor 
    potential~(MTP) that has been successfully deployed to predict 
    structures of a series of aluminium clusters in the range of cluster 
    sizes $N=21-55$~\cite{smith21, wang22}.
    Shiranirad {\em et. al.} have used HDNNP to study the 
    of argon clusters within the cluster size range $N=2-16$ using a 
    training data set of 3483 configurations~\cite{mozh22}.
    The accuracy and reliability of the DeePMD interatomic 
    potential have been tested by Fronzi~\cite{fronzi23}~{\em et. al.} 
    to predict properties of gold nanoparticles within the particle size $N=20-147$.
    A different approach of targetting charge density instead of total energy
    has been proposed and applied successfully to ionic NaCl clusters by
    Godeckar and coworkers~\cite{nacl15}.

    One of the most widely used MLPs that is known to give quantum accuracy 
    is based on the Gaussian Process Regression~(GPR) technique commonly 
    known as the Gaussian Approximation Potential~(GAP).
    This method, introduced by Bart\'ok~\cite{bartok13, schmitz18}~{\em et al.}, 
    requires a relatively small data set.
    It has been widely applied to a variety of systems~\cite{brown19, deringer17, bk19}, 
    demonstrating impressive accuracy and efficiency in predicting atomic 
    forces and energies~\cite{shapeev16}.
    The method has been used successfully to study chemical reactions, 
    material properties, and phase transitions~\cite{bk19, kolb20, artrith16}.
    Sivaraman~\cite{sivraman21}~{\em et al.} have used experimentally 
    produced data sets to develop the GAP potential for HfO$_2$.
    The mean-absolute-error~(MAE) in energies was 2.4~meV/atom.
    The availability of the experimental data was found to significantly 
    reduce the model development time and human efforts.
    Another successful application of the GAP model was to study 
    the $c$-Si/$a$-SiH interface system~\cite{davis22}.
    Here, the model was trained on a data set obtained from MD 
    simulations carried out at different temperatures on the 
    crystalline, amorphous phases of silicon and hydrogen, and their interface systems.
    The resultant GAP model was able to capture not only all the 
    signatures of the DFT results but also accurately reproduce 
    various structural features including partial pair correlation functions, 
    bond angle distributions, etc.
    The model has also been applied successfully to investigate the 
    structural properties of the LiCl-KCl mixture.
    Thus, the above studies clearly indicate the capabilities of 
    the GAP model to accurately predict the properties of various systems.
    In the present work, we have developed interatomic potentials using 
    the GAP model to investigate the structural and thermodynamic 
    properties of sodium clusters in the size range of $55-252$.

    Our choice of free sodium clusters is motivated by a number 
    of attractive and interesting properties of sodium and its clusters.
    Sodium atoms have only one valence electron and therefore 
    larger clusters are easily computable at the DFT level at 
    least for fixed-point total energy calculation.
    Extensive experimental measurements on the melting of sodium clusters 
    in the size range of $N=55-355$ have been reported by Haberland 
    and coworkers~\cite{martin98, haber02, haberland2002}.
    Our earlier works demonstrated the necessity of DFT 
    calculations for understanding the behaviour of melting 
    temperatures~\cite{chk05, ghazi09}.
    However, all the DFT studies were restricted to small 
    sizes~($N \le 142$) \cite{ghazi09, aquado01}, and with 
    limited data. 
    It may be noted that such thermodynamic studies need 
    extensive finite temperatures MD simulations which require 
    accurate force calculations at every step.
    Therefore, computing thermodynamic properties present 
    an ideal playground for testing ML potentials.
    The clusters of sodium offer a class of systems for rigorously 
    testing/comparing the model energies and forces with those 
    of DFT even for larger sizes. 
    
    The remainder of this paper is organized as follows.
    In section~\ref{sec:method}, we describe the methodology and computational
    details. In section~\ref{sec:results}, we present the results 
    on model generation, heat capacities for $N=55$ and 147, 
    as well as geometric isomers of $N=147, 200, 201$ and 252.
    In section~\ref{sec:summary}, we summarize our results along with 
    some important concluding remarks. 

    \section{Methodology and Computational Details}
    \label{sec:method}
    
    There are three essential ingredients required for the development of
    machine learning-based potential energy surfaces.
    The first one is the choice of a suitable machine learning model.
    The second one is the generation of data with sufficient accuracy and the
    third one is the training and validation of the model.
    As noted in the introduction, we have chosen the Gaussian Process 
    Regression technique-based model commonly known as the Gaussian
    Approximation Potential~\cite{bartok10,bartok13}.

    \subsection{The GAP Model}
    We begin by giving a brief summary of the GAP model.
    Following Behler and coworkers~\cite{behler07}, almost all the
    ML models for interatomic potentials use local 
    neighbourhood approximation, where the total 
    energy~($E$) of the system, is given by $E=\sum_{i}E_i$,  
    where $E_i$ is effective atomic energy for the $i^{th}$ atom 
    in the system.
    Thus, it is assumed that the contribution to $i^{th}$ atom 
    sites for the energy and forces come from a neighbourhood 
    within a sphere of radius $r_{cut}$.
    In fact, it is this approximation that enables these models
    to be deployed for larger sizes while training on small systems.
    Another crucial ingredient is the use 
    of set of descriptors required to parameterize the total energy 
    of the system. 
    The descriptors are rotationally, translationally, and 
    permutationally invariant and derived from the coordinates 
    of the neighbourhood atoms for each atom in the cluster.
    A number of large sets of descriptors have been developed by various 
    groups~\cite{pinheiro21, tong20}.
    
    We briefly summarize the Gaussian process regression method. 
    For the sake of completeness, we briefly summarize Gaussian Process 
    Regression method. GPR is a non-linear, non-parametric regression 
    method based on Bayesian probability theory. 
    Consider a smooth function Y({\bf{x}}), where {\bf{x}} multidimensional vector. For a 
    given {\bf{x}}, `Y' maps it  onto a single real scalar value. 
    Dataset consists of a number of samples
    Y$_n$, as a function of {\bf{x$_n$}}. Obviously, the functional form of `Y' is not
    known.
    We approximate Y({\bf x}) by $\bar{Y}$({\bf x}). The 
    function $\bar{Y}$({\bf x}) is approximated by equation,
    \begin{equation}
    \bar{Y}({\bf x})=\sum_{m=1}^{M}c_mk({\bf{x},x_m})
    \end{equation}
    where k({\bf x, x$_m$}) is the basis function located in input space 
    of {\bf x$_m$}.
    In the present case, set `{\bf x}' corresponds to all the coordinates 
    of atoms in the system and `Y' is the total interaction energy.
    The fitting of the model is carried out by minimizing the loss function `$L$', 
    with respect to the coefficient c$_m$, 
    \begin{equation}
    L=\sum_{n=1}^{N}\frac{[Y_n-\bar{Y}({\bf x})]^2}{\sigma_n^2} + R
    \end{equation}
    where `R' is the regression term and $\sigma_n$ is called the 
    relative hyperparameter.
    In practice, the basis function `k' is taken as Gaussian,
    \begin{equation}
    k({\bf x, x_m})=exp({-\frac{\vert{{\bf x-x_m}}\vert^2}{\sigma_{length}^2}})
    \end{equation}
    These basis functions capture the local atomic 
    environment.

    In the GAP implementation, we have used the following descriptors,
    viz., 2-body~({\tt 2b}), 3-body~({\tt 3b}), 
    and smooth overlap of atomic positions~({\tt SOAP})~\cite{bartok13}.
    The {\tt 2b} and {\tt 3b} descriptors are distance and angle-dependent
    respectively.
    For many body interactions, we have used the {\tt SOAP} descriptor.
    The model has a number of hyperparameters which are fixed prior to the process 
    of minimization. However, these parameters have also been optimized by carrying 
    out several minimization runs till the desired accuracy is reached.
    As we shall see, apart from the hyperparameters, the choice of proper 
    selection of data points also
    play a crucial role in the optimization process.

    \subsection{Generation of Data}
    \label{ssec:gen}
    For fitting the GAP model, input data points were generated within the
    the framework of DFT as implemented in the Vienna {\em Ab Initio} Simulation
    Package~(VASP) ~\cite{prb-kre93,prb-kre99}.
    We used force-consistent energy with entropy.
    Several thousand configurations spanning a large configuration
    space of interest is needed.
    Since we are dealing with free clusters, virials are not required.
    In addition, we also need isolated atom energy.
    The generalized gradient approximation~(GGA) with Perdew, Burke, and
    Ernzerhof~(PBE) functional~\cite{prl-per96} has been used for treating
    the exchange-correlation potential.
    All calculations have been performed within the projected augmented
    wavefunction method~(PAW)
    All the configurations were generated using Born-Oppenheimer molecular
    dynamics~(BOMD)~simulations with an energy cut-off of 102~eV on the plane
    wave basis set.
    It is most convenient to generate the configurations by using
    BOMD in a range of temperatures covering the {\em solid-like} 
    to the {\em liquid-like} region in the phase space of each cluster.
    For instance, we generated approximately 10,000 configurations for
    11~temperatures each in a range of 50~K to 400~K for Na$_{55}$ cluster.
    In all the cases, the energy tolerance for the self-consistent field~(SCF)
    calculations was kept at of 10$^{-5}$~eV for each BOMD iteration.
    Note that the purpose of BOMD was to generate the data points
    spanning a sufficient spread in the configuration space.
    This process generated over 100,000 configurations for Na$_{55}$ cluster
    having a total energy range of $\approx~6.0$~eV.
    Two points may be noted.
    First, it is not necessary to generate data points via a molecular dynamical
    simulation.
    Second, a large number of data points is not necessary for an accurate fit.
    Indeed, as we shall demonstrate, we have been able to generate
    a GAP model that can give results with quantum accuracy for a
    range of cluster sizes between 40 to 250 using less than a
    total of 1900 data points.
    This data set includes about 600 data points of Na$_{55}$ cluster augmented
    by small set generated for $N=60, 70, 80, 90, 92, 116, 138, 147, 178$ and
    200.
    It is crucial to mention that these data points spanned the entire
    energy range of interest.

    \subsection{Training the GAP Model}
    To gain insight into the development of GAP models
    and their applicability to a wide range of sizes, we have
    developed two different models using data on single
    and multiple-size clusters respectively.
    A GAP model trained with sufficiently large data points of a single-size
    cluster augmented by smaller data points of other sizes is expected to be valid for
    larger cluster.
    Sodium clusters is homogeneous with just one electron per
    atom.
    Therefore, it was possible to generate a large number of configurations for
    different sizes. We developed two models - the first one trained on data 
    points generated for Na$_{55}$, and the other one uses data points from 
    clusters of other sizes. 
    Now, we discuss the details of the methodology used for training the models.

    The first model (M1) was trained using the data
    set for Na$_{55}$ only.
    Let us recall that one of our objectives is to deploy our model for 
    thermodynamics properties namely heat capacity and melting temperature.
    Therefore, it is desirable to use a training set over a wide range of
    energies.
    The data obtained by BOMD first curated by removing the duplicate 
    configurations that lie within $\Delta{E}$=10$^{-3}$~eV. 
    The resulting configurations $\approx~22,000$ are arranged in an 
    ascending order in energy.
    We first selected about 500~data points that were uniformly distributed
    over the above-mentioned energy range.
    The remaining ones were used for validation of the model.
    We used three classes of descriptors \texttt{2b}, \texttt{3b} and
    \texttt{SOAP} to fit the GAP model.
    Primarily, this data set was used for the optimization of hyperparameters
    such as $r_{cut}$, sparsification scheme $N_{sparse}$, etc.
    In some of the runs, the number of training data points was increased 
    by a few hundred.
\begin{table}\small
\begin{center}
\caption{Final optimized hyperparameters used for 
model M1 and M2. $\Delta{r}$ is transition 
width for SOAP, $\zeta$ denotes the dimensionality of the descriptor, 
$\sigma_{at}$ control the smoothness of the respective kernals, 
$\eta$, and $\it{l}$ are integer indices for orthonormal radial 
basis functions and N$_{sparse}$ is number of sparse points for 
sparse kernal basis.}
\begin{tabular}{c|c|c|c}
\hline
Parameters & \multicolumn{3}{c}{Descriptors}\\ \cline{2-4}
                        & 2-body    &  3-body   &  SOAP  \\\hline
$\delta$                & 0.001     &  0.001    & 0.040  \\
r$_{cut}$ (\AA)         & 12.9      &  3.5      &  12.9  \\
$\Delta{r}$ (\AA)       &           &           &  1.0   \\
$\sigma_{at}$(\AA)      &  0.6      &  0.6      &  0.6   \\
$\eta$                  &  -        &  -        &  12.0  \\
$\it{l}$                &  -        &  -        &  6.0   \\
$\zeta$                 &  -        &  -        &  4.0   \\
Sparsification          & Uniform   &  Uniform  & CUR    \\
N$_{sparse}$            & 20        &  100      &  6000  \\
\hline
\end{tabular}
\label{tab:1}
\end{center}
\end{table}
    Table~\ref{tab:1} gives the final optimized parameters used.
    A high cut-off value of $r_{cut}$ and that of $N_{sparse}$ for the {\tt
    SOAP} descriptor demands significantly large RAM, and 
    become a limiting factor with machines having limited RAM.
    For example, the best results for this model require $\approx~180$~GB of
    memory across all nodes.
    It also may be noted that the RAM requirement increases significantly with
    the size of the clusters as well as the number of training data points.

    The standard error analysis gives a root-mean-square-error~(RMSE) in
    energies and forces as defined by the following equations:
    \begin{equation}
        E_{rmse}= \sqrt{
                    \frac{1}{M}
                    \sum_{k=1}^{M}
                    \Big (
                        E_{DFT}-E_{GAP}
                    \Big )^2
                 }
    \label{one}
    \end{equation}

    \begin{equation}
        F_{rmse}=\sqrt{
                   \frac{1}{M}
                   \sum_{k=1}^{M}
                   \left [
                      \frac{1} {3N^k_{atom}}
                      \sum_{i=1}^{N^k_{atom}}
                      \sum_{\alpha}
                      \Big (
                      F_{k,i,\alpha}^{DFT}-F_{k,i,\alpha}^{GAP}
                      \Big )^2
                   \right ]
                 }
    \label{two}
    \end{equation}
    where, $M$ is the number of configurations, $N^{k}_{natoms}$ is the number
    of atoms in the $k^{th}$ configuration, and $\alpha = x, y, z$ corresponds to the
    three force components of each atom.
    
    \begin{figure}
    \begin{tabular}{ll}
    \includegraphics[scale=0.5]{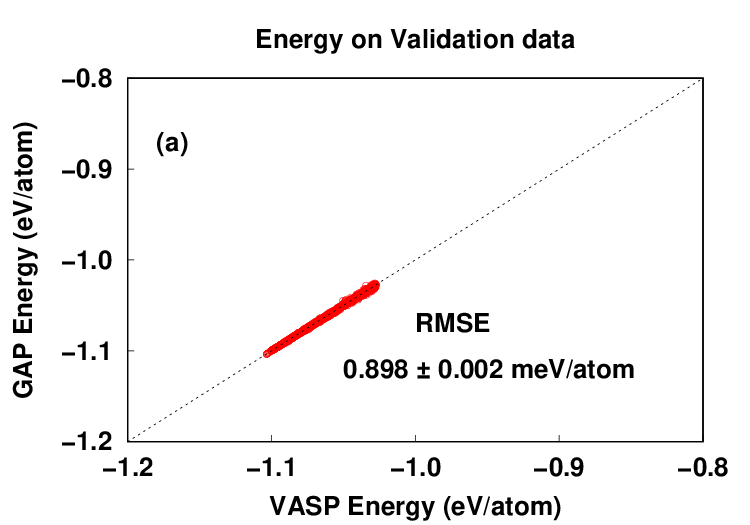}&
    \includegraphics[scale=0.5]{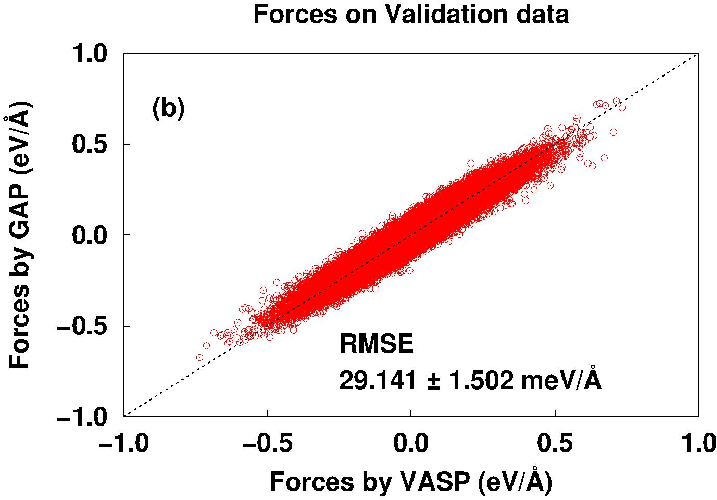} \\
    \includegraphics[scale=0.5]{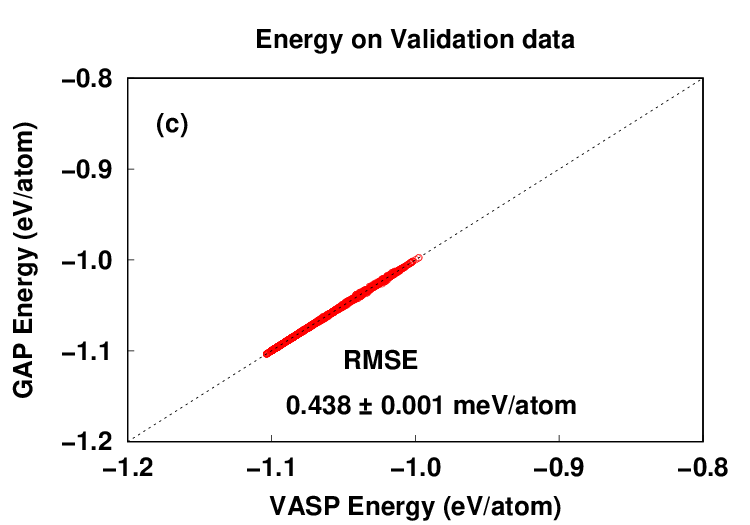}&
    \includegraphics[scale=0.5]{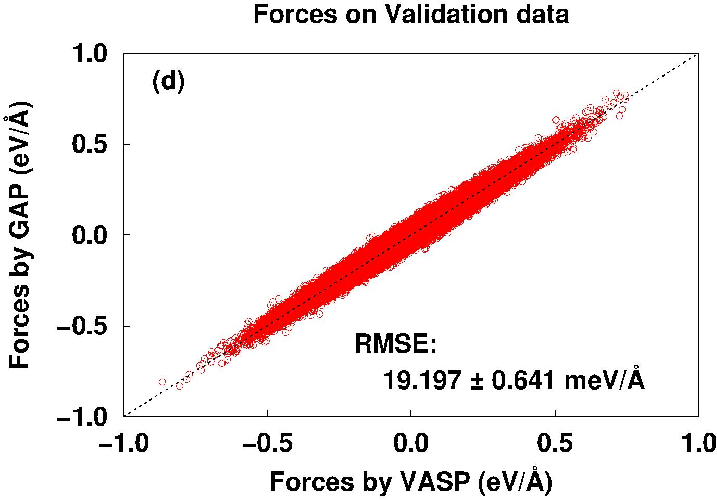} \\
    \end{tabular}
   \caption{Comparison between DFT and GAP-predicted energy and forces for Model M1.
            Figure~\ref{fig:1}(a) and (b) corresponds to M1-1 and Figure~\ref{fig:1}(c) 
	    and (d) corresponds to M1-3.}
    \label{fig:1}
    \end{figure}

    It may be noted that the standard RMS values are averages. 
    Since, we wish to carry out long MD simulations at temperature
    slightly above the melting temperatures, the errors in forces can 
    become critical.
    Therefore, after examining the errors in the forces of individual 
    atom along with errors in the energies, we augmented the training 
    data set by adding the configurations near the points having absolute 
    force above a certain value, leading to an increase of 100-150 data 
    points in the training data set.
    In the present case, the errors turn out to be uniformily distributed (see Figure~\ref{fig:1}). 
    Therefore, it was convenient to increase the number of data points uniformily over the entire energy range. 
    The models were retrained, and this process was carried out iteratively 
    until the maximum error in the forces was less than 0.1\%.
    With this procedure, the final size of the training data set 
    was about 1900 configurations. This is designated as model M1.

    In Figure~\ref{fig:1}, we show the comparison between DFT quantities and GAP results of 
    energies and forces for validation data. Figure~\ref{fig:1}(a) and Figure 1(b) 
    correspond to M1-1 and Figure~\ref{fig:1}(c) and Figure 1(d) correspond to M1-3.
    It can be seen that there is dramatic improvement in the accuracy of the model. 
    The RMS values for energy and forces for final model M1-3 are 0.438~meV/atom 
    and 19.197~meV/$\AA$ respectively. As shown in the Table~\ref{tab:2}, the maximum 
    error in the force is of the order of 0.1$\%$. In fact, we have examined the
    absolute error in energy and forces for the all the validation configurations. 
    These plots are shown in supplementary information S5 and S6. 

\begin{table}[!ht]
	\caption{Maximum percentage errors in energy and forces 
	during training of Na$_{55}$. Out of several training sessions,
	three representative ones are shown. The maximum of all 
	forces (3~$\times$~number of configuration~$\times$~N) is noted.}
    \centering
    \small\begin{tabular}{c|c|c|c|c|c|c}\hline
        Model &\multicolumn{2}{c|}{Configurations}&\multicolumn{4}{c}{Maximum percentage Error}\\\cline{2-7}
              & Training & Validation & Training & Validation & Training & Validation \\ \hline
        M1-1 & 544  & 1899 & 0.532 & 0.938& 0.509 & 0.562 \\ 
        M1-2 & 630  & 1101 & 0.033 & 0.353& 0.096 & 0.108 \\ 
        M1-3 & 1945 & 4184 & 0.024 & 0.311& 0.113 & 0.129 \\ \hline
    \end{tabular}
    \label{tab:2}
\end{table}

    The model M1 was successfully used in computing the specific heat for Na$_{55}$.
    However, when the model was employed to test larger systems
    up to $N=250$, in some cases the errors in the energies and
    the percentage error in the forces increased up to about 32~meV and 
    40~$\%$, respectively.
    This impelled us to improve on model M1 by including data points 
    from other cluster sizes.
    Typically, we included 20-50 data points for each cluster of
    sizes N=40-55, 60, 70, 80, 90, 92, 116, 138,
    147, 178, and 200.
    The resultant model is designated model M2.
    The comparison between model M1 and M2 is given in Table~\ref{tab:3}. 
    It is gratifying to see that the absolute errors in the energies 
    reduce significantly to less than 0.94 meV. We also examine the 
    maximum force for all the case and is found to less than 0.11$\%$. 
    (see supplementary information S7 and S8.)

\begin{table}[ht]                                                                                                  
 \caption{Maximum absolute error in energy, Mean absolute error~(MAE) and 
	root mean square error~(RMSE) for validation sets of various sodium clusters. 
	The unit of energy is given in meV/atom.}
     \centering
   \small\begin{tabular}{c|c|c|c|c|c|c|c}\hline
   System & No. of Configs & \multicolumn{6}{c}{Error in Energy (meV/atom)}
   \\\cline{3-8}
 &&\multicolumn{2}{c|}{Absolute}&\multicolumn{2}{c|}{MAE}&\multicolumn{2}{c}{RMSE}\\\cline{3-8}
 &&M1 & M2& M1&M2& M1&M2\\\hline
%
70	& 2541  &28.729 &	4.76 &24.563 &0.828&24.614 &1.062  \\ 
80	& 6178  &35.098 &	5.08 &30.477 &0.833&30.556 &1.092  \\
90	& 1355  &49.752 &	2.79 &44.868 &0.644&44.960 &0.849  \\
92	& 2045  &60.888 &	2.87 &51.500 &0.876&51.621 &1.127  \\
116	& 3556  &59.536 &	1.61 &56.576 &0.391&56.593 &0.488  \\
138	& 4324  &90.624 &	0.94 &87.214 &0.140&87.223 &0.191  \\
147	& 3672  &82.865 &	1.16 &80.778 &0.170&80.784 &0.229  \\
178	& 1372  &91.929 &	1.23 &89.866 &0.236&89.872 &0.346  \\
200	& 2148  &99.451 &	1.53 &95.815 &0.217&95.821 &0.296  \\\hline
 \end{tabular}
     \label{tab:3}
 \end{table}

    For this final model M2, 1213 data points were used of which only
    50 data points per cluster were used from the large size clusters.
    It is of some interest to point out that for the case of large cluster e.g. Na$_{200}$, 
    each data point generates about 200~neighbourhoods or 200~finite-sized clusters which is 
    nearly 4 times larger than those generated by the 55-atom cluster.
    Thus, adding only a few data points of clusters with a larger number
    of atoms, improves the quality of the model significantly~(see supplementary
    information S9).
    Indeed, model M2 (the final best model) required a RAM of more than 650~GB.

    \section{Results and Discussion}
    \label{sec:results}

    We now employ these models to obtain the thermodynamics
    properties, specifically heat capacities of Na$_{55}$ and Na$_{147}$, 
    as well as to obtain the isomers of Na$_{147}$, Na$_{200}$, Na$_{201}$, 
    and Na$_{252}$ clusters. We start with a discussion of the 
    thermodynamic properties.

    \subsection{Thermodynamics}

    We have used the well-established and well-tested multiple histogram 
    methods for computing specific heat. For the multiple histogram method 
    density of accessible states~$\Omega(V, T)$, are required for 
    a large range of temperatures.
    Care is taken to ensure that there is sufficient overlap between 
    the adjacent histograms.
    This essentially decides the choice of temperature intervals.
    For more details of the multiple histogram method, the reader is 
    referred to the articles~\cite{ferren90, dgk}.
    We have carried out molecular dynamics simulations using
    the LAMMPS package~\cite{plimpton95} with GAP models M1 and M2.
    We have computed the heat capacities for Na$_{55}$ and Na$_{147}$.

\begin{figure}
    \begin{tabular}{ll}
    \includegraphics[scale=0.55]{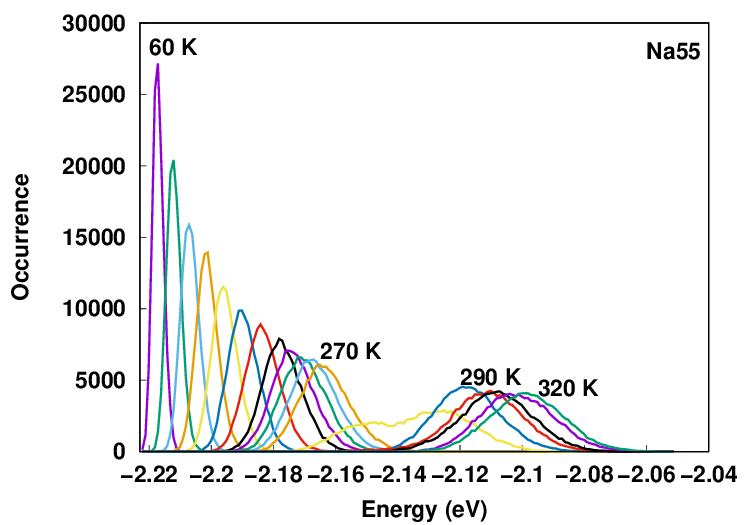}&
    \includegraphics[scale=0.55]{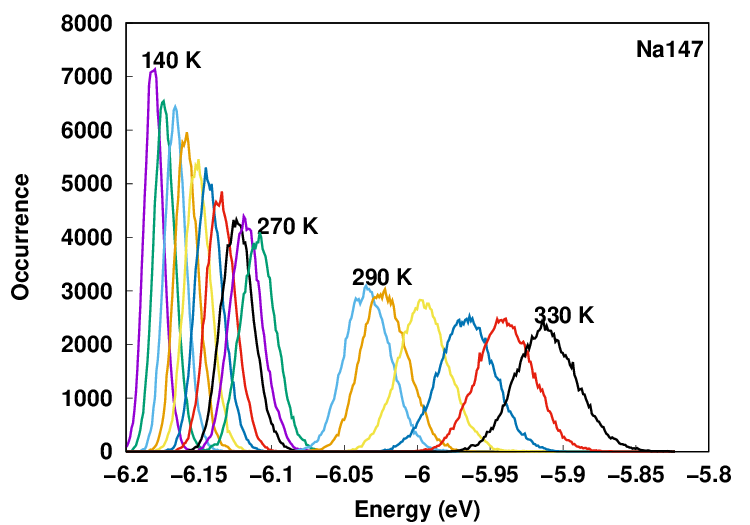} \\
    \end{tabular}
    \caption{The multiple histograms showing the potential energy
       distribution for Na$_{55}$ and Na$_{147}$. At higher
       temperatures the histograms are much wider, indicating
       liquid-like behaviour.}
    \label{fig:2}
    \end{figure}

    The multiple histograms are generated by performing MD simulations
    at 21 different temperatures ranging from 60~K to 330~K for both
    clusters, with a time step of ~2~fs using Nos\'e-Hoover
    thermostat leading to a total simulation time of 400~ps per temperature.
    All simulations began with the lowest energy structure as obtained
    by our models.
    We have used the GAP model M1 for Na$_{55}$ and model M2 for
    Na$_{147}$.
    The lowest energy structures for Na$_{55}$ and Na$_{147}$ are
    icosahedra.
    The ground state structure of Na$_{147}$ is shown in
    Figure~\ref{fig:ground}.
    It agrees very well with our earlier density-functional
    result~\cite{ghazi09}.
    Figure~\ref{fig:ground} also depicts the lowest energy 
    structures of Na$_{200}$, Na$_{201}$ and Na$_{252}$ clusters.
    The multiple histogram obtained for Na$_{55}$ and Na$_{147}$ are 
    shown in Figure~\ref{fig:2}.
    A clear separation of peaks between two temperatures 270~K 
    and 290~K indicates the temperature range of melting.
    This feature should also be reflected in a peak in the heat 
    capacity plot.

    \begin{figure}
    \begin{tabular}{ll}
    \includegraphics[scale=0.55]{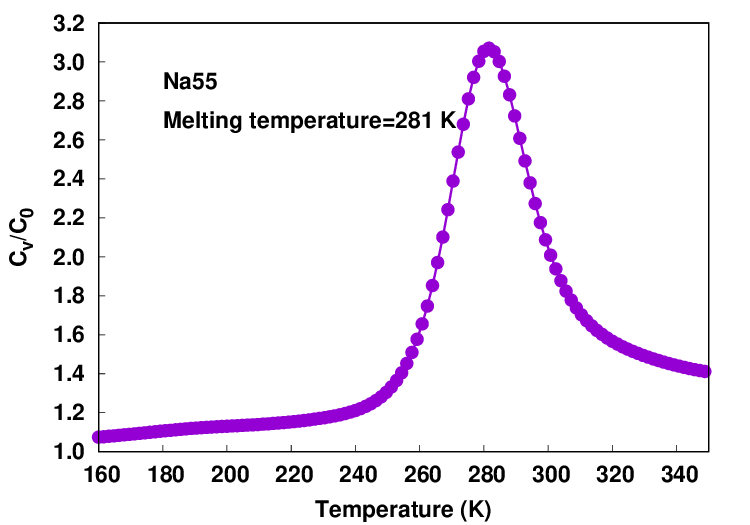}&
    \includegraphics[scale=0.55]{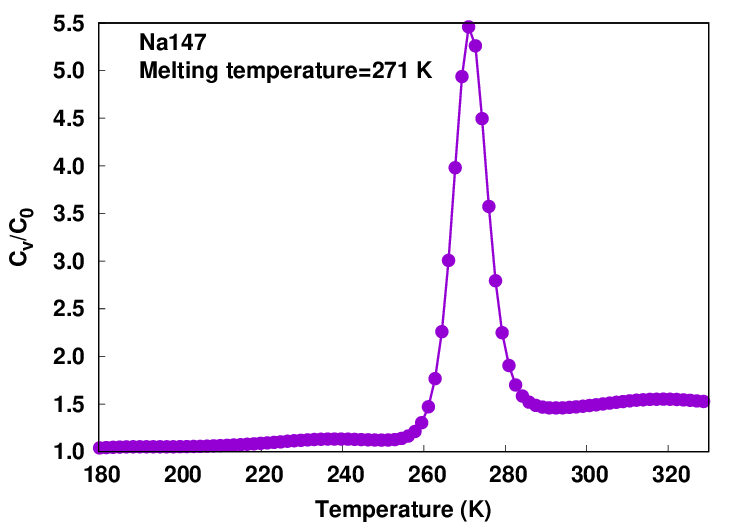}\\
    \end{tabular}
    \caption{Heat capacity curves for Na$_{55}$ to
    Na$_{147}$ using MD simulation with GAP model. The heat capacity curves are
    scaled C$_0$, where C$_0$=(3N-9/2)k$_B$ is the zero temperature classical 
    limit of the rotational plus vibrational canonical specific heat.}
    \label{fig:3}
    \end{figure}

    The resulting heat capacities $C_v/C_0$, where C$_0$=(3N-9/2)k$_B$ is 
    the zero temperature classical limit of the rotational plus 
    vibrational canonical specific heat as a function
    of temperature $T$ for Na$_{55}$ (model M1) and
    Na$_{147}$ (model M2) clusters are shown in Figure~\ref{fig:3}.
    It can be observed that the melting temperatures identified by 
    the peaks in the heat capacities are 281~K ($\pm 10$~K)
    and 271~K~($\pm 10$~K), for Na$_{55}$ and Na$_{147}$, respectively.
    There is an excellent agreement between the melting temperatures 
    obtained in this work with those measured 
    experimentally by Haberland {\em et. al.} as well
    as with earlier reported density-functional calculations~\cite{chk05}.
    Note that the reported experimental values by Haberland {\em et. al.}
    are 290~K and 272~K for Na$_{55}$ and Na$_{147}$, respectively.
    This is the first reported result on specific heat calculations of
    Na$_{147}$ using machine learning potential with density-functional accuracy.

    The above discussion clearly shows that the GAP models we developed
    are accurate enough for sodium clusters.
    As we shall see, the models can be extended to obtain
    isomers and even thermodynamics of clusters of sizes for $N=200, 201$ 
    and 252 sodium atoms. It may be noted that the time taken by the 
    GAP model to carry out the MD simulation over that of 
    the {\em ab initio} methods is tremendously decreased~\cite{testrun}.
    We have exploited this fact to obtain 40 isomers each 
    for $N=147, 200, 201$ and 252.

    \subsection{Isomers using GAP Model M2}
    We now turn our attention to obtaining isomers of clusters
    of sizes of more than 150 atoms. Specifically, we present results
    for 40 isomers of sodium clusters of sizes $N=147, 200, 201$, and $252$.
    It may be noted that the previous density-functional studies on sodium 
    clusters are mainly restricted to $\leq$ 150~atoms~\cite{ghazi09, aquado01}, 
    most of them being below 100 atoms~\cite{manna23}.

\begin{table}[!ht]
~\hspace{-2.0cm}
\caption{Percentage error in energies and bond length of few lowest energy isomers generated using GAP model M2.
$n$ is the isomer number.}
{\scriptsize
\begin{tabular}{c|c|c|c|c|c|c|c|c|c|c|c|c|c}\hline
& \multicolumn{6}{c|}{Na$_{147}$}&\multicolumn{6}{c}{Na$_{200}$}\\\cline{2-7}\cline{3-14}
        ~ & \multicolumn{3}{c|}{Energy (eV)} & \multicolumn{3}{c|}{Bondlength (\AA)} & \multicolumn{3}{c|}{Energy (eV)} & \multicolumn{3}{c}{Bondlength (\AA)}\\\hline
        $n$  & VASP & GAP    & Error & VASP & GAP & Error &	$n$  & VASP & GAP    & Error & VASP & GAP & Error \\ \hline
        1 & -170.6213  & -170.9273 & 0.1793 & 3.493 & 3.494 & 0.0001 &    1 & -234.8059 & -235.1887& 0.1630& 3.366 & 3.360 & 0.165 \\ 
        2 & -170.4499  & -170.7899 & 0.1994 & 3.441 & 3.265 & 5.110 &    2 & -234.2854 & -234.6999& 0.1769& 3.336 & 3.289 & 1.399 \\ 
        3 & -170.3532  & -170.6974 & 0.2020 & 3.306 & 3.263 & 1.324 &    3 & -234.3508 & -235.2440& 0.3811& 3.339 & 3.315 & 0.732 \\ 
        4 & -170.1236  & -170.6238 & 0.2940 & 3.263 & 3.260 & 0.091 &    4 & -233.2936 & -234.4981& 0.5163& 3.280 & 3.276 & 0.137 \\ 
        5 & -169.3874  & -170.1951 & 0.4768 & 3.260 & 3.267 & 0.236 &    5 & -233.2315 & -234.3009& 0.4584& 3.292 & 3.289 & 0.098 \\ 
        6 & -168.7203  & -170.0023 & 0.7598 & 3.267 & 3.212 & 1.693 &    6 & -231.7990 & -233.3538& 0.6707& 3.313 & 3.284 & 0.883 \\ 
        7 & -168.6236  & -169.8945 & 0.7536 & 3.230 & 3.235 & 0.157 &    7 & -232.4981 & -233.6663& 0.5024& 3.255 & 3.278 & 0.722 \\ 
        8 & -168.5279  & -169.8162 & 0.7644 & 3.247 & 3.230 & 0.527 &    8 & -232.0359 & -233.4958& 0.6291& 3.279 & 3.292 & 0.396 \\ 
        9 & -168.3992  & -169.4272 & 0.6104 & 3.248 & 3.273 & 0.785 &    9 & -232.2099 & -233.3486& 0.4903& 3.311 & 3.294 & 0.527 \\ 
        10& -167.7377 &  -169.1585 & 0.8470 & 3.285 & 3.274 & 0.350 &   10 & -232.3758 & -233.4144& 0.4469& 3.226 & 3.204 & 0.665 \\ \hline
   & \multicolumn{6}{c|}{Na$_{201}$}&\multicolumn{6}{c}{Na$_{252}$}\\\cline{2-7}\cline{3-14}
           ~ & \multicolumn{3}{c|}{Energy (eV)} & \multicolumn{3}{c|}{Bondlength (\AA)} & \multicolumn{3}{c|}{Energy (eV)} & \multicolumn{3}{c}{Bondlength (\AA)}\\\hline
           $n$  & VASP & GAP    & Error & VASP & GAP & Error & $n$  & VASP & GAP    & Error & VASP & GAP & Error \\ \hline
        1 & -235.2917 & -236.8940 & 0.6810& 3.269 & 3.410 & 4.308 &   1 & -296.6961 & -298.5249 & 0.6164 & 3.332 & 3.363 & 0.924 \\ 
        2 & -235.2835 & -236.8564 & 0.6685& 3.316 & 3.437 & 3.621 &   2 & -296.6267 & -298.5029 & 0.6324 & 3.370 & 3.379 & 0.264 \\ 
        3 & -235.2445 & -236.7578 & 0.6432& 3.250 & 3.390 & 4.310 &   3 & -296.6022 & -298.4893 & 0.6362 & 3.413 & 3.379 & 0.991 \\ 
        4 & -235.1744 & -236.6899 & 0.6444& 3.231 & 3.286 & 1.698 &   4 & -296.6018 & -298.4621 & 0.6272 & 3.383 & 3.270 & 3.336 \\ 
        5 & -235.1206 & -236.4845 & 0.5800& 3.261 & 3.393 & 4.029 &   5 & -296.5856 & -298.4434 & 0.6263 & 3.398 & 3.341 & 1.677 \\ 
        6 & -234.9514 & -236.4387 & 0.6330& 3.236 & 3.289 & 1.620 &   6 & -296.5463 & -298.4230 & 0.6328 & 3.402 & 3.337 & 1.886 \\ 
        7 & -233.5629 & -235.5166 & 0.8364& 3.260 & 3.285 & 0.752 &   7 & -296.5218 & -298.4002 & 0.6335 & 3.383 & 3.342 & 1.198 \\ 
        8 & -233.3826 & -235.4676 & 0.8933& 3.185 & 3.293 & 3.383 &   8 & -296.4960 & -298.3887 & 0.6383 & 3.386 & 3.329 & 1.702 \\ 
        9 & -233.3317 & -235.4161 & 0.8933& 3.273 & 3.293 & 0.590 &   9 & -296.4904 & -298.3555 & 0.6290 & 3.364 & 3.326 & 1.114 \\ 
       10 & -233.3260 & -235.3186 & 0.8539& 3.293 & 3.273 & 0.586 &  10 & -296.4846 & -298.3180 & 0.6183 & 3.291 & 3.292 & 0.014 \\\hline 
\end{tabular}
}
\label{tab:4}
\end{table}

    First, let's discuss the method of obtaining the isomeric structures.
    We have carried out extensive MD simulations of the order of 400~ps
    for each cluster at six temperatures in the range of 100~K-300~K.
    This gave a total of 1.2 million configurations, from which we
    choose every 100$^{th}$ configurations for local minimization.
    The configurations were then sorted according to their energies,
    and only one in the set of all degenerate configurations that lay
    within $\Delta{E}~\le~0.01eV$ were selected.
    This led to about 200 structures that were used for local minimization 
    using the M2 model. These 200 structures were further scrutinized 
    for any duplicates giving only about 40 distinct structures per cluster.

\begin{figure}
     \includegraphics[width=1.0\textwidth]{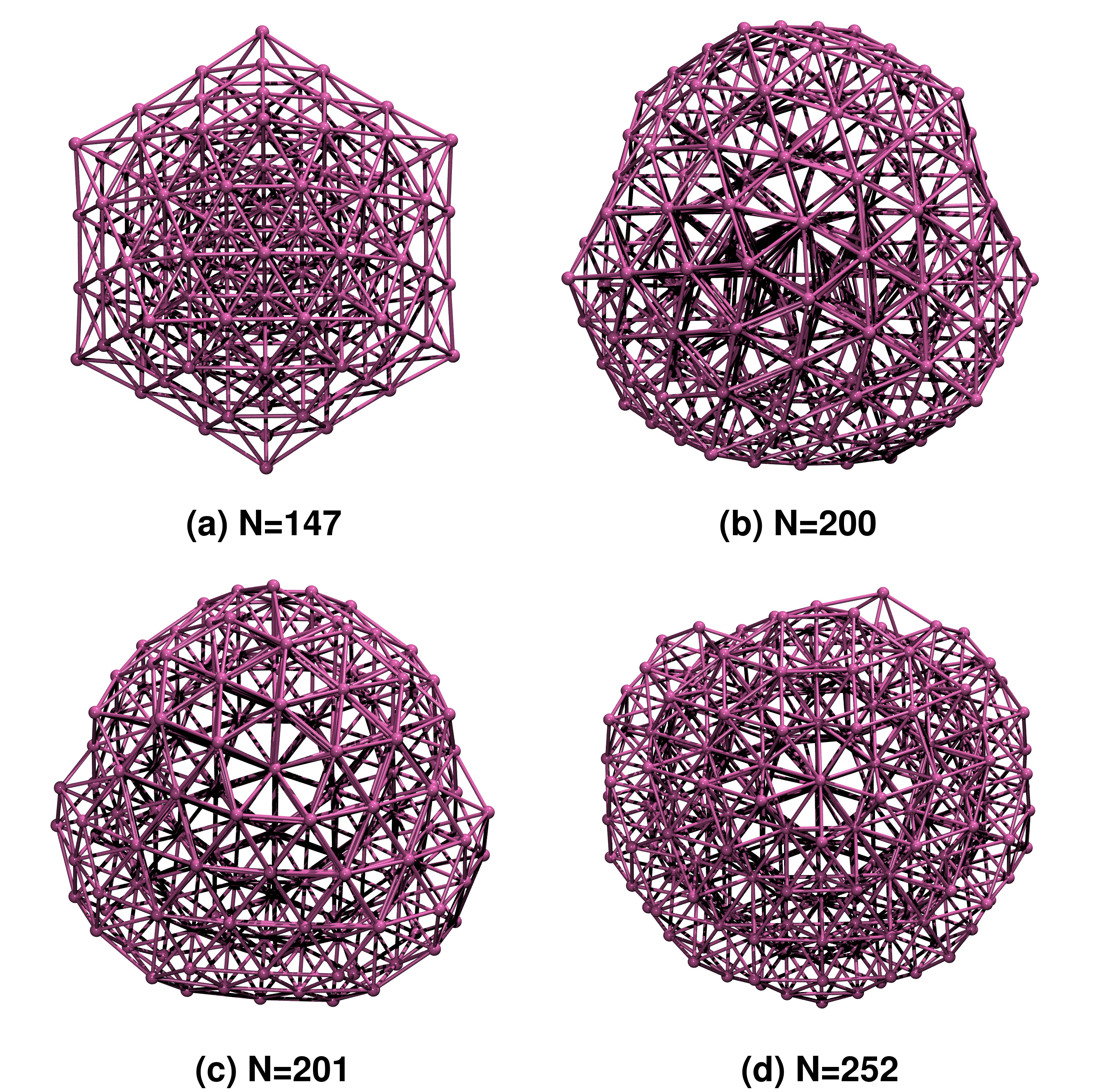}
     \caption{Ground state structures of Na$_{147}$, Na$_{200}$,
                 Na$_{147}$, and Na$_{252}$ sodium clusters.}
     \label{fig:ground}
     \end{figure}

    To assess the accuracy of the GAP model, the 20 lowest energy 
    structures were further optimized using DFT.
    A comparison between the total energies and bond lengths, 
    giving percentage error for the 10 lowest structures are presented 
    in Table~\ref{fig:ground}. In all the cases the error in energy is 
    less than 0.9\%. While the errors in bondlength for Na$_{147}$ and 
    Na$_{200}$ are less than 1\%~(with four exceptions). The errors are 
    slightly more for Na$_{201}$. The mean absolute error 
    ($|E_{DFT}~-~E_{GAP}|$) is less than 7~meV/atom, the smallest 
    and largest errors being 1.9~meV/atom and 10.3~meV/atom respectively.
    Thus, it can be observed that geometries obtained with the GAP 
    models are in good agreement with those computed by DFT. It may be 
    emphasize that only a few iterations by VASP were required 
    to get equilibrium structures.

    \begin{figure}
    \includegraphics[width=1.0\textwidth]{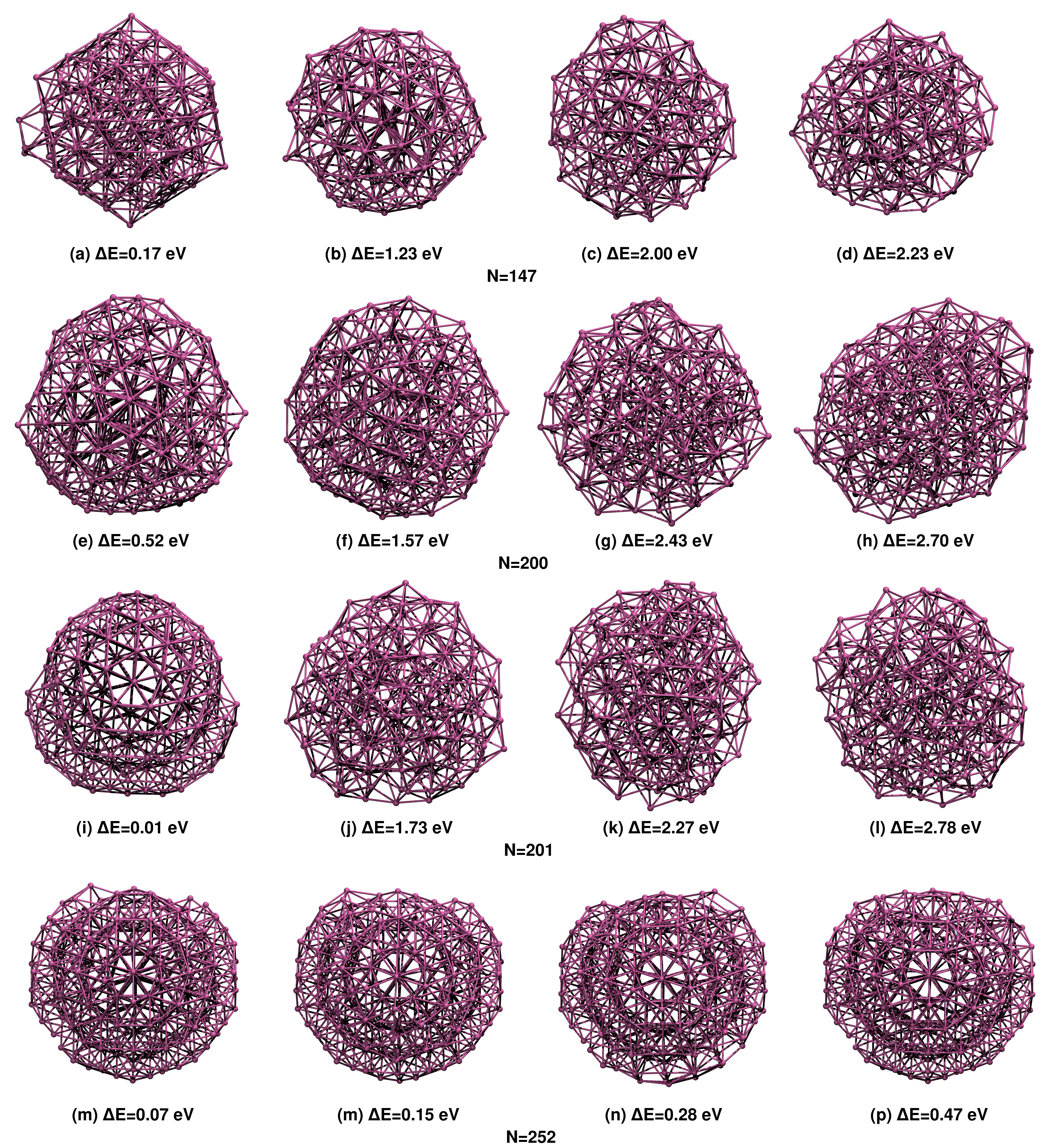}
    \caption{Lowest-energy structures of Na$_{147}$, Na$_{200}$,
		Na$_{147}$, and Na$_{252}$ sodium clusters.}
    \label{fig:isomers}
    \end{figure}

    In Figure~\ref{fig:ground}, the lowest energy geometries for Na$_{147}$, 
    Na$_{200}$, Na$_{201}$, and Na$_{252}$ are shown. 
    Some representative geometries of low and high energy isomers 
    are shown in Figure~\ref{fig:isomers}.  
    We can see from Figure~\ref{fig:ground}, the lowest-energy
    structure of Na$_{147}$ is complete icosahedron.
    The shapes of the 200 and 201 atoms clusters can be
    understood as distorted icosahedron surrounded by
    the remaining atoms, forming a partial third shell on one side.
    This gives these geometries a non-spherical shape.
    Interestingly, addition of extra atom to Na$_{200}$, changes 
    the shape significantly. In Na$_{201}$, third shell is far 
    more spherical than Na$_{200}$. From Figure~\ref{fig:ground}, 
    we see significant differences in the structural features of 
    the Na$_{200}$ and Na$_{201}$ clusters. The geometries of 
    both clusters are quite distorted. The 252-atom sodium cluster is 
    also distorted, and yet very different from those of 200 and 
    201-atom clusters. Here, the shape is nearly spherical.

    Figure~\ref{fig:isomers} shows some of the high-energy structures 
    of sodium clusters. 
    In each of the cases, the first isomer shown is closest in energy to 
    the gound state. The remaining high energy structures
    are randomly selected. It is observed that the high-energy 
    structures are considerably distorted.  
    The computing the ground state and equilibrium geometries are non-trivial, 
    especially for for large clusters. A systematic understanding of evolution of 
    geometries, ordering of isomers and nature of high energy isomers is 
    beyond the scope of present work. In this work, we have developed machine 
    learned model which is accurate enough to address some of these problems.

    \section{Summary and Conclusion}
    \label{sec:summary}
    We have developed machine-learned Gaussian Approximation
    Potential models for constructing the potential energy surfaces
    for sodium clusters for a wide range of sizes.
    We demonstrate that about 1300 data points are sufficient to 
    yield very accurate models.
    These models have been used to compute the thermodynamic properties 
    of Na$_{55}$ and Na$_{147}$ using multiple histogram method. The 
    results of the model are in excellent agreement with experimental 
    heat capacity curves as well as DFT results.
    The models has enabled us to compute the heat capacity curve and melting
    temperature of Na$_{147}$ which has not been previously reported. 
    In addition, up to 20 isomers of the larger cluster ($N\ge200$)
    that were difficult to obtain with the accuracy of DFT have been 
    reported with density-functional accuracy.
    Global optimization of cluster geometry of large sizes is known to be highly compute 
    intensive. Our work demonstrates that ML model paves a way for obtaining 
    several isomers of large clusters of sodium.
    Even without additional minimization by DFT, used in the present work, 
    the total energy and geometries are fairly accurate.
    Having established the accuracy of GAP models, it is tempting 
    to employ the models to explore several
    properties of much larger clusters which were not accessible
    with limited computational resources.
    We hope the availability of such models will open up a way to investigate
    physics of large clusters.

   \section{Acknowledgements}
    One of us (DGK) is pleased to acknowledge a number of useful 
    discussions with Dr. Balchandra Pujari, Centre for Modeling 
    and Simulation, SPPU, Pune and Dr Sandip De, BASF, Germany. 

    \section{CRediT authorship contribution statement}
    All results have been obtained by equal contributions of the authors.

    \section{Declaration of competing interest}
    The authors declare that they have no known competing 
    financial interests or personal relationships that could have
    appeared to influence the work reported in this paper.

    \section{Data Availability}
    The data used in training and validation as well as model parameters have 
    been provided in the supplementary information. In addition, the coordinates of
    isomer geometries obtained with GAP and VASP are also available.

    \bibliographystyle{unsrt}
    \bibliography{bibliography}

\begin{thebibliography}{10}
\bibitem{testrun}
In a test simulation for 10,000 MD steps with a timestep of 3~fs we got a
  scaling of about 300 times by LAMMPS over VASP.
\end{thebibliography}

    \end{document}